\documentclass[useAMS,usenatbib]{mn2e}
\usepackage{mhchem} %chemformula
\usepackage{lscape}
\usepackage{graphicx}
\title{2D condensation model for the inner Solar Nebula: an enstatite-rich environment}
\author[Pignatale et al.]{F. C. Pignatale$^{1,2}$\thanks{E-mail:
francesco.pignatale@univ-lyon1.fr}, Kurt Liffman$^{2}$, Sarah T. Maddison$^{2}$, Geoffrey Brooks$^{3}$\\
$^{1}$ 	Universit\'e de Lyon, Lyon, F-69003, France; Universit\'e Lyon 1, Observatoire de Lyon, 9 avenue Charles Andr\'e, Saint-Genis
Laval, F-69230, France; \\ CNRS, UMR 5574, Centre de Recherche Astrophysique de Lyon; Ecole Normale Sup\'erieure de Lyon,
F-69007, France \\
$^{2}$Centre for Astrophysics and Supercomputing, Swinburne
University of Technology, Hawthorn, VIC 3122, Australia\\
$^{3}$FSET, Swinburne University  of Technology, Hawthorn, VIC 3122, Australia}

\begin{document}
\renewcommand{\thefigure}{\arabic{figure}}

\date{{Accepted in MNRAS on 2015 December 23. Received 2015 December 16; in original form 2015 May 04.}}

\pagerange{\pageref{firstpage}--\pageref{lastpage}} \pubyear{2002}

\maketitle

\label{firstpage}

\begin{abstract}
Infrared observations provide the dust composition in the protoplanetary discs surface layers, but can not probe the dust chemistry in the midplane, where planet formation occurs. Meteorites show that dynamics was important in determining the dust distribution in the Solar Nebula and needs to be considered if we are to understand the global chemistry in discs. 1D radial condensation sequences can only simulate one disc layer at a time and cannot describe the global chemistry or the complexity of meteorites. To address these limitations, we compute for the first time the two dimensional distribution of condensates in the inner Solar Nebula using a thermodynamic equilibrium model, and derive timescales for vertical settling and radial migration of  dust.

We find two enstatite-rich zones within 1~AU from the young Sun: a band $\sim$0.1~AU thick in the upper optically-thin layer of the disc interior to 0.8~AU, and in the optically-thick disc midplane out to $\sim$0.4 AU. The two enstatite-rich zones support recent evidence that Mercury and enstatite chondrites shared a bulk material with similar composition. Our results are also consistent with infrared observation of protoplanetary disc which show emission of enstatite-rich dust in the inner surface of discs. 

The resulting chemistry and dynamics suggests that the formation of the bulk material of enstatite chondrites occurred in the inner surface layer of the disc, within 0.4~AU. We also propose a simple alternative scenario in which gas fractionation and vertical settling of the condensates lead to an enstatite-chondritic bulk material.
\end{abstract}


\section{Introduction}
\label{sec-intro}

Infrared spectroscopy probes the chemistry of the surface layers of protoplanetary discs, but  provides little information about the dust composition of layers deep inside the disc and the midplane
\citep{2004ASPC..309..213C,2011ppcd.book..114H}.  Furthermore, the derived chemistry from the upper layers of discs is unlikely to reflect the dust composition of the midplane, since the physical conditions of these regions may be very different. Thus, information about the bulk chemistry of  the disc, where the planet formation process takes place, is missing and hence modelling is required.

One dimensional radial condensation sequences, which resemble the midplane of the Solar Nebula with an initial solar gas mixture, provide general agreement with the derived bulk chemical composition of the Solar System planets \citep{1995GeCoA..59.3413Y,1998A&A...332.1099G}, but they cannot provide more detailed insights such as the complex chemistry of meteorites, and the location in which their bulk composition formed. Furthermore, 1D condensation sequences can simulate only one layer of the disc at time and cannot account for the global chemistry of discs.

The dust in protoplanetary disc is subject to a series of dynamical processes \citep{2011ARA&A..49..195A}. In particular, vertical and radial transport of grains clearly played a role in the early Solar System in determining the dust distribution throughout the disc \citep{2005A&A...443..185B}. The behaviour of the dust due to aerodynamic drag is determined by the stopping time, $t_s = ({\rho_d s_d })/({c_s \rho_g})$, where $\rho_d$ and $s_d$ are the intrinsic density and the size of the dust grains and $c_s$ and $\rho_g$ are the gas sound speed and the gas density at a given location \citep{1977MNRAS.180...57W,2005A&A...443..185B}. Thus, the decoupling of the dust grains is regulated by its aerodynamic parameter, $\zeta=\rho_d s_d$  \citep{2006mess.book..353C}: denser and/or larger grain have longer stopping times than smaller and/or lighter grains.

The effect of dynamic processes on the composition and distribution of the dust in discs becomes clear when the morphology of meteorites is analysed. Chondrites are characterized by heterogeneous compositions \citep[]{2005mcp..book..663B} which show mixtures of compounds and features (calcium-aluminium-rich inclusion, chondrules, etc.) resulting from several processes (condensation, aqueous alteration and metamorphism) that occurred  in different environments and at different times during the protoplanetary disc phase of our Solar System \citep[]{2005mcp..book..143S}. The processes which aggregated high and low temperature materials are still unconstrained. Moreover, analysis of rare objects like enstatite chondrites suggest that very unique chemical conditions must have been present in the early Solar Nebula \citep{2012Weisberg}. To date, there is no general consensus on the chemical pathways which generated the enstatite chondrites or the location in the Solar Nebula in which they formed.

In order to address these limitations and obtain more information on the processes that occurred in the early Solar Nebula, a new approach in the study of the chemistry of protoplanetary dust is needed. Detailed disc models have been developed in recent years which incorporate thermal and dynamical processes to study their structure and evolution \citep[]{2010A&A...513A..79B,2008PhST..130a4015D,1999ApJ...527..893D,1998ApJ...500..411D}. These disc models provide the two dimensional temperature and pressure distribution within each zone of the disc, from the midplane to the surface, and can be used to couple physical, dynamical and chemical studies of gas and dust in discs. Furthermore, the availability of  chemical software packages and computational resources allow us to study complex systems which could not be investigated before.

We present a new study which attempts to link the observed chemistry in discs, and the chemical evidence derived from analysis of meteorites. Our aims are to (i) provide insights into the possible origin of the crystalline silicates observed in protoplanetary discs, (ii) determine the locations in which the bulk of rare objects, such as enstatite chondrites, might have formed, and (iii) provide insight into the chemical composition of the layers deep inside the disc which cannot be probed by infrared observations.

 In this work we utilise a 2D disc model to derive, for the first time, the  condensates distribution within the Solar Nebula, mainly focusing on the main silicates observed in discs: forsterite (\ce{Mg2SiO4}) and enstatite (\ce{MgSiO3}). The resulting distribution is compared with observations of protoplanetary discs and combined with analytical calculations of dynamical processes (radial migration and vertical settling of dust and the extension of the dead zone).

The paper is ordered as follows: we describe our method and its limitations in section~\ref{sec-method}, as well as present our fiducial disc model which represents our initial condition, describe the thermodynamic model used and the analytical treatment of dynamics. Our results are presented in section~\ref{sec-results}. In section~\ref{sec-EHmodel} we investigate the possible locations of formation of the enstatite chondrites' bulk material, and suggest mechanisms of formation. In section~\ref{sec-discussion} we discuss the theoretical and observable consequences of our results, and link infrared disc observations with the bulk material of enstatite chondrites and the dust in the Solar Nebula midplane.  Our conclusions are presented in section~\ref{sec-conclusions}.

\section{Method}
\label{sec-method}
In this section we describe the properties of our fiducial disc model,  the thermodynamic equilibrium model used to calculate the condensate distribution, and the analytical approach we use to solve for the vertical settling and radial migration of the dust in the disc and the location of the dead zone.

Before proceeding, we stress the limitations of our thermodynamic calculations. In the low temperature regions of the disc, kinetic barriers will prevent some chemical processes from occurring. Therefore, the true composition of the dust in these zones can diverge from the compositions predicted when assuming complete equilibrium. Given the degree of complexity of the systems we consider, this work only focuses on the thermodynamic aspect in order to build the necessary framework for future work, which could investigate kinetics and test the reliability of the proposed scenarios. In the following sections we stress the regions of the discs for which  equilibrium is a valid assumption and where it is not.

\subsection{Disc model}
\label{subsec-disco}
The temperature and pressure distribution within the disc, $T(R,Z)$ and $P(R,Z)$, are determined using the 2D disc model of \citet{1998ApJ...500..411D,1999ApJ...527..893D}. Heating  sources in the disc include viscous dissipation and radioactive decay, which can generate heat at each location of the disc. Also, there are cosmic rays and stellar irradiation, which penetrate the disc from the surface and interact with the gas and dust. The effects of disc irradiation by accretion shocks on the stellar surface are also included. We chose stellar parameters to mimic the young Sun, with $M_{\ast}=1 \, {\rm M}_{\odot}$, $R_{\ast}=2.6 \, {\rm R}_{\odot}$,  $T_{\ast}=4278$~K,  $L_{\ast}=2.069~L_{\odot}$, $\dot{M}=10^{-8} \, {\rm M}_{\odot} {\rm yr}^{-1}$, and viscosity parameter $\alpha=0.01$, assuming a 1~Myr old star with isochrones calculated from \citet{1997A&A...324..556S,2000A&A...358..593S}. At this relatively late evolutionary phase, the accretion rate is low and the central star has already accreted most of the material which will constitute its final mass. We therefore  assume that the remaining dust in the disc will constitute the building blocks of larger objects. 1~Myr is a reasonable time to overcome all the kinetic barriers which can affect dust formation (condensation and/or annealing), especially in the inner regions of the disc where the temperature is high enough, and  we can use this evolutionary phase to consider equilibrium.

\begin{figure*}
{\includegraphics[width=1.\columnwidth]{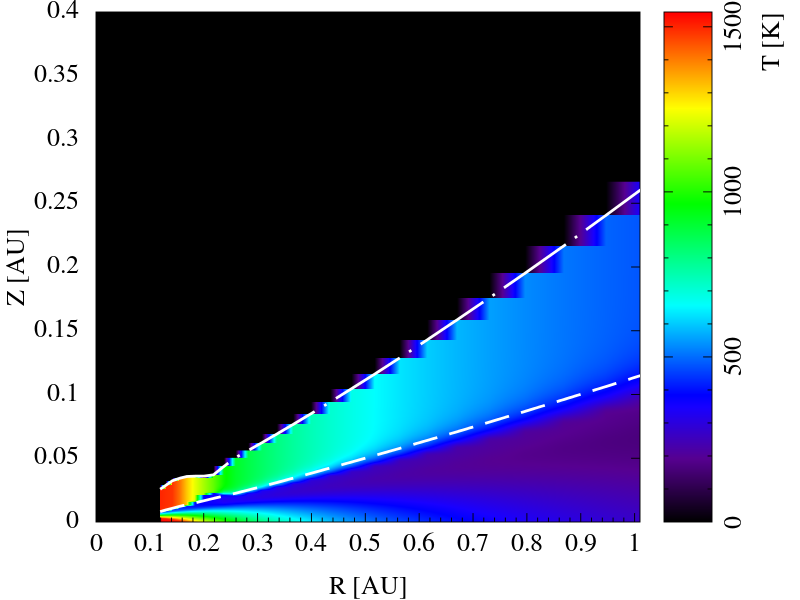}}
{\includegraphics[width=1.\columnwidth]{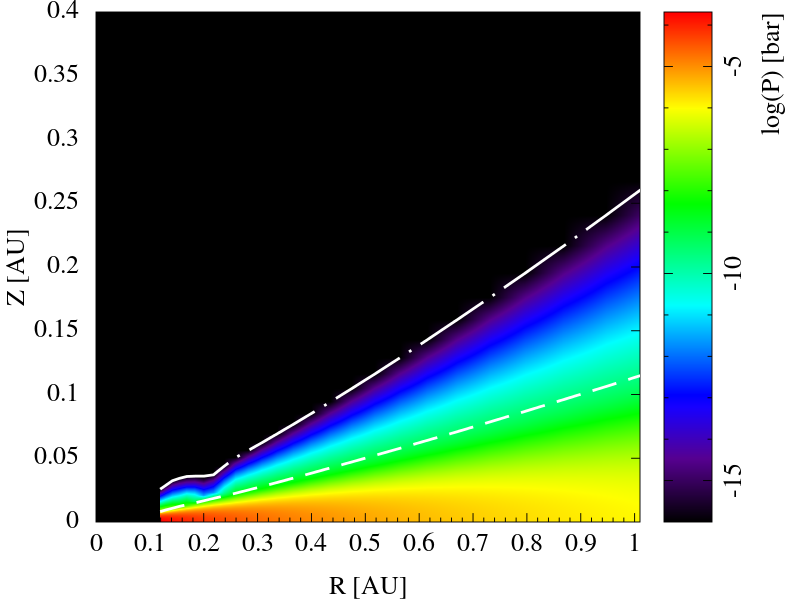}} \\
\caption{Resulting disc structure from the \citet{1998ApJ...500..411D} model. Left: temperature distribution. Right: pressure distribution. The dashed line is the $\tau=1$ surface of the disc and the dashed-dot line represents the disc surface. }
\label{fig-tandp}
\end{figure*}

The temperature range in the disc spans $50\le T\rm{(K)}\le 1450$ and the pressure ranges  $10^{-16}\le P\rm{(bar)}\le10^{-4}$. In Fig.~\ref{fig-tandp} we present the disc structure and also show the $\tau=1$ surface under which the disc becomes optically thick \citep{1998ApJ...500..411D,1999ApJ...527..893D}. The disk ``surface'' is defined by the edge of the \citet{1998ApJ...500..411D,1999ApJ...527..893D} grid for the disc model we are using. In the midplane, the stellar radiation is strong enough to heat the disc to over 1000~K. Thus,  equilibrium can be a reasonable assumption for the optically thin surface layer out to 0.8~AU and in the midplane out to 0.4~AU. The optically thick zone of the disc beyond 0.4~AU, where the temperature decreases dramatically, will not be considered in our discussions.

\subsection{Thermodynamic model}
\label{subsec-thermomodel}
 
We derive the 2D distribution of condensates by determining the thermodynamic equilibrium of an initial gas mixture, given a set of temperatures and pressures, using the Gibbs free energy minimisation technique \citep[]{DeHoff1993}. 

We utilise the FactSage software package \citep[]{Bale2002,Bale2009}, which uses the minimisation method described by \citet{Eriksson1990} and \citet{Eriksson1995}. The thermodynamic data for each compound are taken from the database provided by FactSage\footnote{http://www.crct.polymtl.ca/fact/documentation/}. The initial gas mixture is composed of the 15 most abundant elements of the solar photosphere from \citet{2009ARA&A..47..481A}, with their abundances normalised to 100~kmol (see Table~\ref{tab-abundance},  column S). We assume that the gas is initially homogeneous throughout the disc and we perform equilibrium calculations using ($T,P$) at each location ($R,Z$) in the disc. The list of possible compounds that can condense comprise 170 gases and 317 solids.

We use the ideal solution for modelling the phase behaviour in this region of the disc, which is widely used in astrophysics to compute  the chemical bulk material which characterise  discs and exoplanets \citep{2005Icar..175....1P,2010ApJ...715.1050B,2012ApJ...759L..40M}. It is known that the ideal solution model is often a poor approximation of the phase behaviour especially in the low temperature regime.  However, the choice of the ideal solution model was made for several reasons: (i) at high temperatures, which characterise the zone of the disc in which we focus our research, the solution behaviour of phases approaches the ideal \citep{kaptay2012}; (ii) there is no significant change, in macroscopic scales, of the condensation sequence and condensed amount for the chemical compounds presented in this work when different solution models are applied \citep{2011MNRAS.414.2386P}; (iii) solution models become important when micro-scales systems are considered, and (iv) the number of  phases in our simulations is large.

\subsection{Dust dynamics}
\label{subsec-dynamics}

The standard accretion disc model  provides a detailed thermodynamic structure of protoplanetary discs \citep{1998ApJ...500..411D,1999ApJ...527..893D}. However, protoplanetary discs are evolving and dynamic objects \citep{2011ARA&A..49..195A}. Dust grains will likely be transported from the location in which they formed to different environments and these motions will likely affect their chemistry. Dust grains which condense out from the gas at different temperatures, will have different intrinsic densities  (compositions) and, as a consequence, different responses to aerodynamic drag. Furthermore, as they grow in size, their stopping time changes. Growing grains can decouple from the gas and settle towards the midplane of the disc \citep{1977MNRAS.180...57W,2005A&A...443..185B,2008A&A...487..265L}.

Thus, in order to understand the evolution of the chemical content of the disc, it is also crucial to consider the effects of different dynamical processes on the dust distribution. In this work we focus on three main processes: the vertical settling and the radial migration of the dust, and the effect of the dead zone on the dust motion.

\subsubsection{Dust vertical settling and radial migration}
 There is a large body of work in the literature which investigates the dynamics of dust in discs:  \citet{2008A&A...487..205D} studied  dust sedimentation from the surface in a turbulent disc and its effect on the resulting 10~$\mu$m infrared spectra, \citet{2010A&A...513A..79B}  also included in their model the effect of gas drag, radial drift  and turbulence in the study of the evolution of the dust growth and dust and gas mixing processes, while \citet{2008A&A...480..859B} investigated the effects of dust coagulation (sticking) and fragmentation on the radial drift of dust particles. 3D models have been developed as well: \citet{2008A&A...487..265L} used a two-phase (gas+dust) Smoothed Particles Hydrodynamics code to investigate the vertical settling and radial migration of growing dust grains within protoplanetary discs. All these studies show that dust mixing is an important process in the redistribution of gas and dust within discs.

In this work, we follow the approach of \citet{1996cppd.proc..285L} and \citet{2000Icar..143..106L} to derive the timescale of dust vertical settling and radial migration within our disc model. 
 The vertical dust settling timescale, $\tau_{set}$, is given by

\begin{equation}
\label{eqn-vertical}
\tau_{set}= 125,000 \frac{ \left(\frac{\rho_{g}}{10^{-11}~\rm{g~cm^{-3}}}\right) \left(\frac{{\it v_{t}}}{1~\rm{km~s^{-1}}}\right) \left(\frac{{\it R}}{1~\rm{AU}}\right)^{3} } {\left(\frac{M_{\star}}{M_{\odot}}\right)\left(\frac{a_{p}}{0.1 \rm{\mu m}}\right)\left(\frac{\rho_{p}}{1 \rm{g~cm^{-3}}}\right)}~\rm{yr} \,, 
\end{equation}
where $\rho_{g}$ is the gas mass density, $v_{t}$ is the Maxwellian speed of the gas, $a_{p}$ is the particle  radius, $\rho_{p}$ is the particle density and $R$ is the radial distance of the dust particle from the star.

Since we aim to derive the magnitude of the dynamic timescales, we approximate the radial velocity of migrating particle via $dR/dt\approx -3\nu/(2R)$ \citep{hartmann2000accretion}, where $t$ is the time and $\nu$ is the average kinematic viscosity.  The migration timescale for a particle at distance $R$ from the star is obtained by integrating this equation, which gives
\begin{equation}
\tau_{mig}=\Delta t= \frac{R^{2}-R_{in}^{2}}{3\nu}~\rm{yr}\,,
\label{eqn-radialmig}
\end{equation}
where $R_{in}$ is the inner boundary of the disc, which we set to 0.1~AU. Since in this work we are focusing on silicate grains, for these calculations the value of the dust density, $\rho_{p}$,  is chosen to represent the average density of silicates: 3 $\rm{g~cm^{-3}}$. 

\subsubsection{MRI and the dead zone}
The Magneto-Rotational Instability (MRI) is thought to be an efficient source of the viscosity which drives accretion in discs \citep{1991ApJ...376..214B}. The {\it dead zone} is a consequence of the MRI: when the ionisation processes drop below a critical value, the gas will not be coupled to the magnetic field. \citet{1996ApJ...457..355G} studied the efficiency of ionisation processes in a layered disc model and found a zone in the midplane, contained between two magnetically active layers, for which neither local ionisation nor thermal ionisation are efficient at driving MRI. This zone would likely be a magnetically inactive zone in which turbulence is suppressed. The extent of this zone has been subject of numerous studies \citep[]{2008MNRAS.388.1223S,2012MNRAS.424.1977L,2012MNRAS.420.3139M} which show that the size of the dead zone depends on several factors such as the number density of the electrons and the chemistry and the size of the dust  \citep{2007prpl.conf..555D}.  In this work, to solve for the extent of the dead zone in the disc, we follow the procedures described by \citet{1996ApJ...457..355G} and \citet{1998ApJ...500..411D}.

We calculate the magnetic Reynolds number, $R_{eM}$, at each location in the disc  following  \citet{1996ApJ...457..355G}:
 \begin{equation}
R_{eM} = 7.4\times10^{13} x \alpha^{1/2} \Bigl(\frac{R}{1 \rm{AU}} \Bigl)^{3/2} \Bigl( \frac{T}{500 \rm{K}} \Bigr) \Bigl( \frac{M_{\star}}{M_{\odot}} \Bigr)^{-1/2} \,,
\label{eqn-reynolds}
\end{equation}
where $\alpha$ is the accretion parameter, $x$ is the ionisation fraction, and $T$ is the local temperature. The MRI, which drives accretion in the disc, will be suppressed if   $R_{eM}\le1$  \citep{1996ApJ...457..355G}. 

Following \citet{1998ApJ...500..411D}, we consider energetic particles, $x_{ep}$,  and thermal ionisation, $x_{th}$, as the main ionisation sources \citep{1992Icar...97..130S,1996ApJ...457..355G,1998ApJ...500..411D,2012MNRAS.420.3139M}. 
The local ionisation parameter is  described by \citet{1992Icar...97..130S} as
\begin{eqnarray}
 \label{eqn-step1}
x_{ep} &=& 5.2\times10^{-18} a_{p}^{-1}T^{1/2} \times \nonumber \\ 
&\times & \Bigl( \sqrt{1+1.0\times10^{-17} \frac{r_{gr}^{2}(\chi_{R}+\chi_{CR})}{\rho^{2}T}} -1\Bigr),
\end{eqnarray}
where $\chi_{R}$  is the rate of ionisation by a radioactive element, given by $\chi_{R}= \lambda_{d} NEn_{H}/36.3$, and  $\chi_{CR}$  the rate of ionisation by cosmic rays, given by $\chi_{CR}=10^{-17}n_{H}e^{-\Sigma/\Sigma_{0}}$,

and $\lambda_{d}$  the decay constant, $N$ the abundance relative to hydrogen, $E$ the average energy available for ionisation (and 36.3 eV is the average energy needed to produce an ion pair in \ce{H2} gas), $n_{H}$ is the hydrogen abundance, and $\Sigma$ is the disc surface density.

The thermal ionisation parameter is described by \citet{2012MNRAS.420.3139M} as
\begin{eqnarray}
\label{eqn-thermal}
x_{th} &=&  6.47\times10^{-13} (10^{-7})^{1/2} \Bigl(\frac{T}{10^{3} \rm{K}}\Bigr)^{3/4} \times \nonumber \\
&\times & \Bigl( \frac{2.4\times10^{15}}{n_{n}}\Bigr)^{1/2}  \frac{e^{-25188/T}}{1.15\times10^{-11}} \,,
\end{eqnarray}
where $n_n$ is the total number density given by $n_{n}= n_{c} \exp (- 0.5 (z/H_p)^{2})$, with $z$  the height above the midplane, $H_p$  the disc scale height and $n_{c}$  the electron number density at the mid-plane given by $n_{c}= \Sigma/(\sqrt{2\pi}\mu m_{H}H_p)$, where $\mu$ is the mean molecular weight and $m_{H}$ the atomic mass of hydrogen.

We set the parameter values as follows:  $a_{p}=0.1~\mu$m, $\rho=\rho$(R,Z)~g~cm$^{-3}$ and $\Sigma=\Sigma$(R) g~cm$^{-2}$  \citep{1998ApJ...500..411D}, $\lambda_d=3\times10^{-14}$~s$^{-1}$, $N=10^{-10}$, $E=3.16$~MeV and  $\Sigma_{0}=100	$ g~cm$^{-2}$  \citep{1992Icar...97..130S},  $n_{H}=10^{13}$~cm$^{-3}$ \citep{1996ApJ...457..355G},   $\mu=2.3$, and  $m_{H}=1.007$ amu \citep{2012MNRAS.420.3139M}. The maximum of $(x_{ep},x_{th})$  is used as input for the total ionisation fraction $x$, for the computation of the magnetic Reynolds number $R_{eM}$ in equation~(\ref{eqn-reynolds}). The choice of the  maximum of $(x_{ep},x_{th})$ is justified by the difference between the two ionisation factors, which can be three orders of magnitude \citep{1998ApJ...500..411D}.

%-----------------------------------------------------------------------
%-----------------------------------------------------------------------

\section{Results}
\label{sec-results}

\begin{figure*}
\centering
{\includegraphics[width=1\columnwidth]{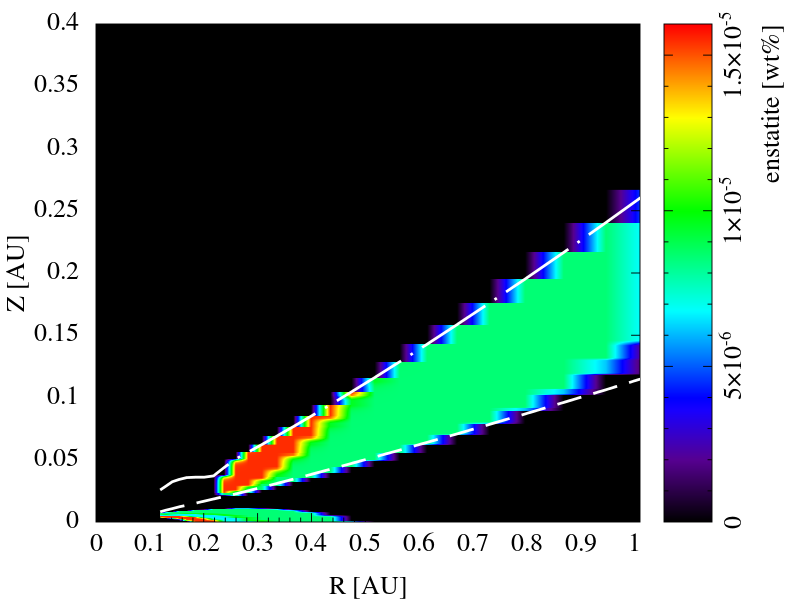}} 
{\includegraphics[width=1\columnwidth]{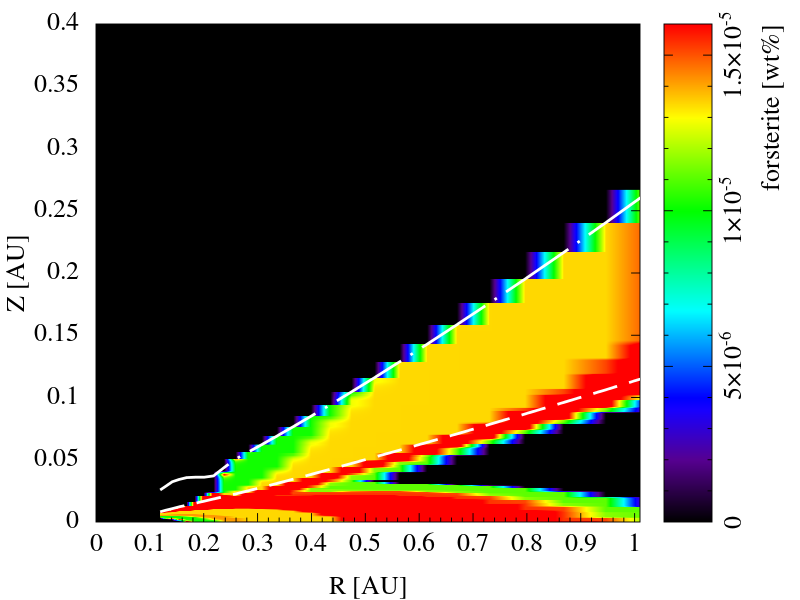}} \\
{\includegraphics[width=1\columnwidth]{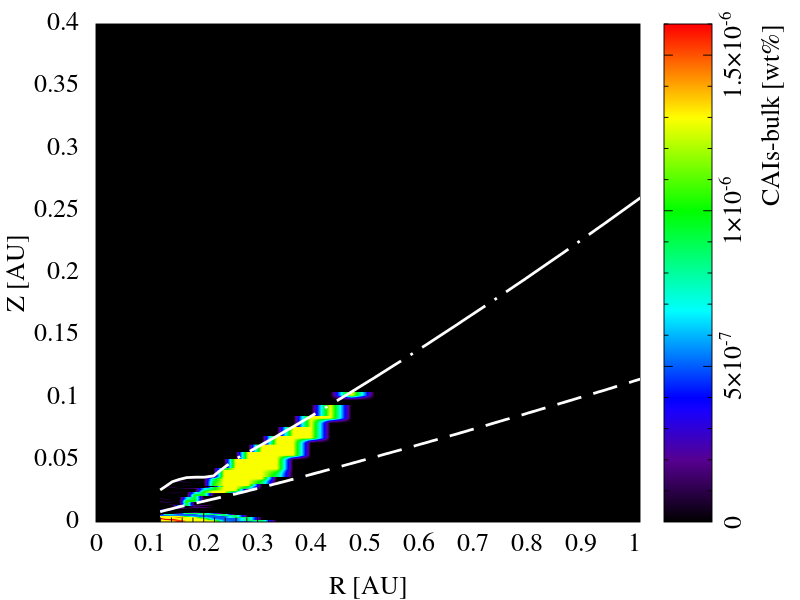}} 
{\includegraphics[width=1\columnwidth]{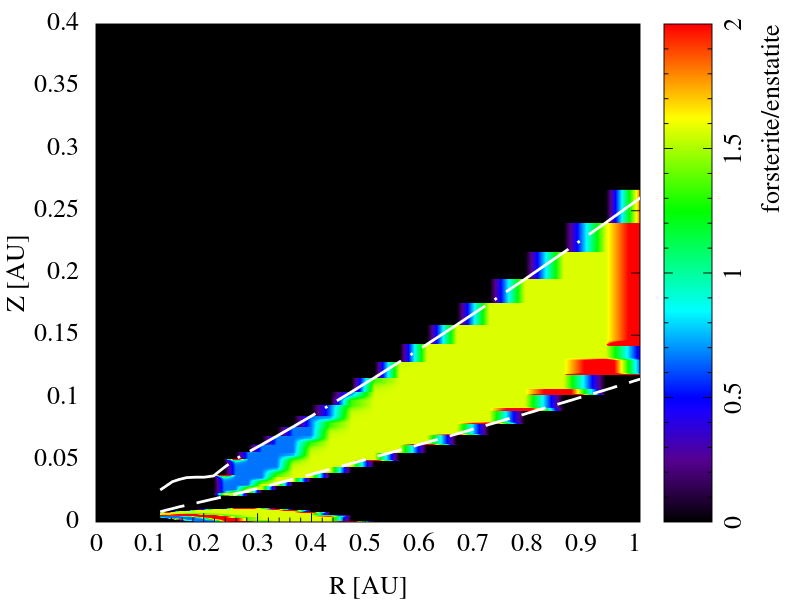}} \\
{\includegraphics[width=1\columnwidth]{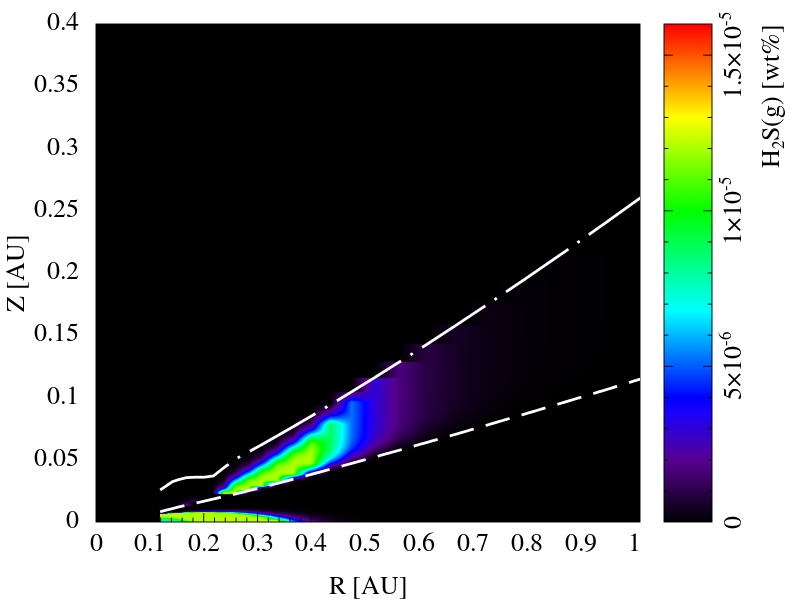}} 
{\includegraphics[width=1\columnwidth]{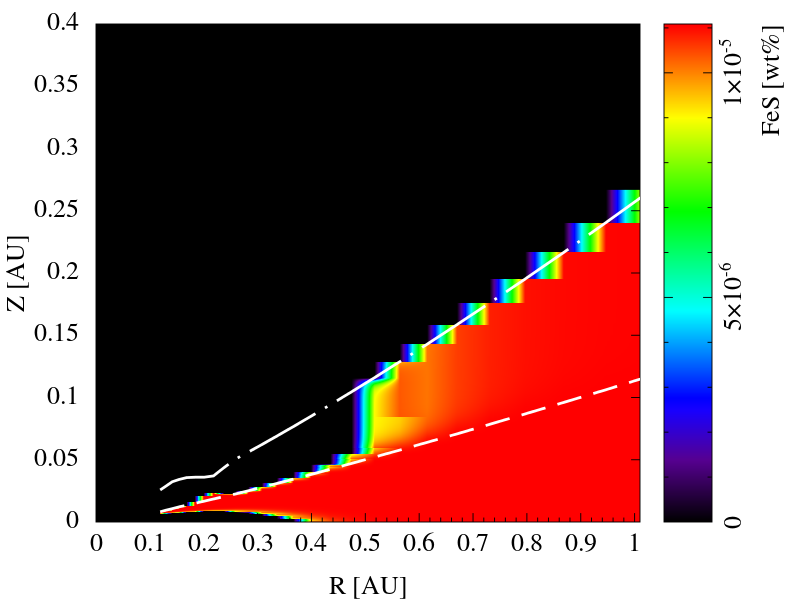}} \\
\caption{Resulting chemical distribution from our 2D model. Top row: enstatite and forsterite distribution. Mid row: CAI bulk components and forsterite-to-enstatite ($fo/en$) ratio. Bottom row:  \ce{H2S}(g) and \ce{FeS} distribution. Note the change of scales. The dashed line is the $\tau=1$ surface of the disc and the dashed-dot line represents the disc surface.}
\label{fig-results}
\end{figure*}

In Fig.~\ref{fig-results} we show the resulting 2D chemical distribution from our model. The top row shows the distribution of  enstatite (\ce{MgSiO3}) and forsterite (\ce{Mg2SiO4}); the middle row shows the location where the calcium-aluminium bulk components condense (or CAI bulk components)\footnote{These include hibonite (\ce{CaAl12O19}),  gehlenite (\ce{Ca2Al2SiO7}),  akermanite (\ce{Ca2MgSi2O7}),   Mg-spinel (\ce{MgAl2O4}), grossite (\ce{CaAl4O7}) and anorthite (\ce{CaAl2Si2O8}).} and the forsterite-to-enstatite  ($fo/en$) ratio; and the bottom row shows the \ce{H2S}(g) and \ce{FeS} distribution.

We find that  stable  enstatite is limited to two well-defined zones within the disc: a band $\sim$0.1 AU thick  in the upper layer of the disc interior to 1 AU, and in the disc midplane out to $\sim$0.4 AU. Forsterite is more abundant in a wider zone in the outer upper layer of the disc and the stability zone in the midplane reaches out to 1~AU. However, as previously stated, the region beyond 0.4~AU falls in the non-equilibrium zone.

There are also two stability zones in the inner 1~AU of the disc where CAI bulk components are present: one in the upper layer of the disc between $0.2\le \rm{R(AU)} \le0.5$ and another 0.01~AU thick zone in the midplane  between the inner boundary of the disc and 0.3~AU. A ``cloud'' of  \ce{H2S}(g) is stable between 0.25 and 0.5~AU below the surface of the disc, and a thick zone of stability out to 0.4~AU is present in the midplane. However, \ce{H2S}(g) can be stable also towards lower temperatures since the formation of \ce{FeS} via \ce{H2S}(g) can be inhibited by kinetic barriers and the effective presence of metallic Fe (see section~\ref{connecting}).

Regions in which $fo/en\le1$ are present in both the surface and midplane of the disc. The average grain composition where $fo/en\le1$ is forsterite (\ce{Mg2SiO4}) 20.2~wt\%, enstatite (\ce{MgSiO3}) 29.6~wt\%, metals (\ce{Fe}-\ce{Ni}) 47.2~wt\%, diopside (\ce{CaMgSi2O6}) 1.3~wt\%, others 1.7~wt\%. No sulfides condensed in this region. 
On the surface of the disc, the main silicates are all distributed in the optically thin region (top row of Fig.~\ref{fig-results}).

\begin{figure*}
\centering
{\includegraphics[width=1\columnwidth]{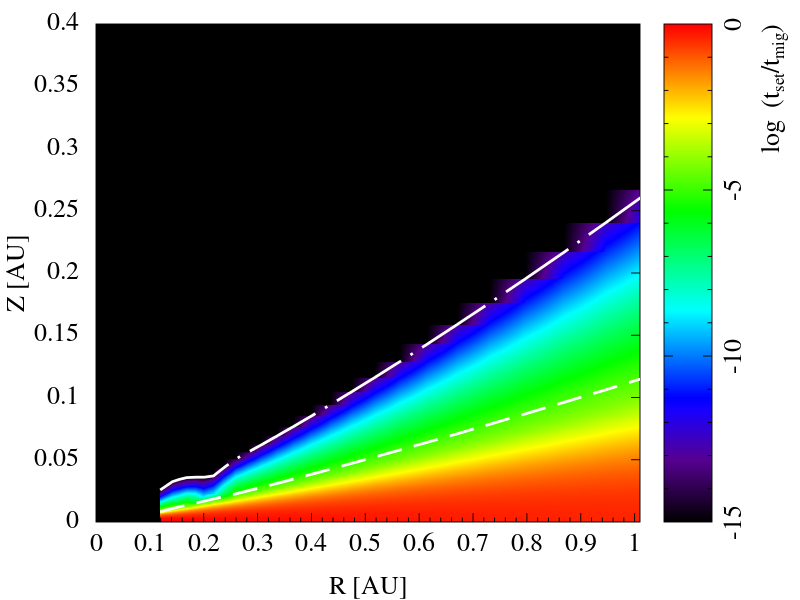}} 
{\includegraphics[width=1\columnwidth]{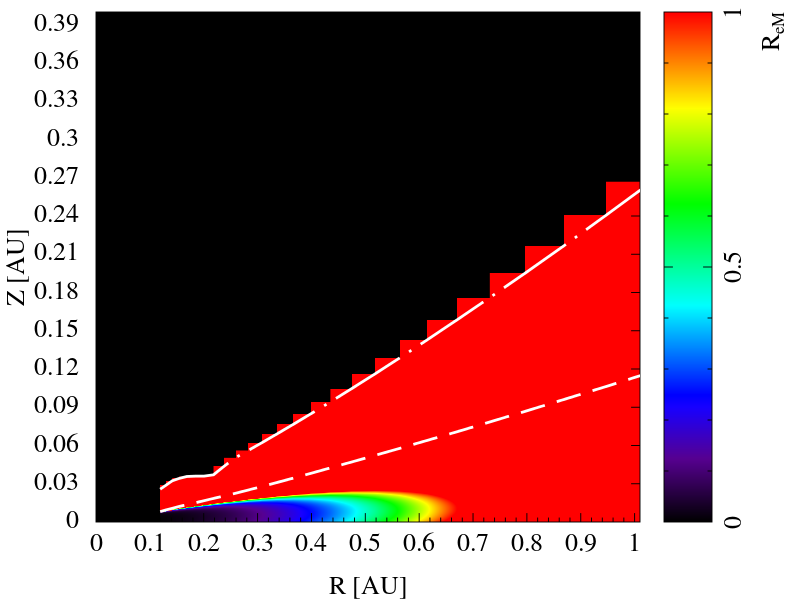}} 
\caption{Dynamical distribution maps: left, $\log$($\tau_{set}/\tau_{mig}$), and right, $R_{eM}$  for 0.1~$\mu$m-sized grain. The dashed line is the $\tau=1$ surface of the disc and the dashed-dot line represents the disc surface.}
\label{fig-timescale02}
\end{figure*}

In Fig.~\ref{fig-timescale02} we present the main dynamical results in our model, showing the distribution of  $\log$($\tau_{set}/\tau_{mig}$) and the extension of the dead zone for 0.1~$\mu$m-sized grains. This grain size is the typical size of  forsterite grains  as derived by  infrared spectral modelling  \citep{2008ApJ...683..479B}. In the upper layers of the disc the settling timescale, $\tau_{set}$, is  shorter than the radial migration timescale, $\tau_{mig}$, and hence grains formed in the upper layers of the disc will generally  settle to the midplane before radially migrating. The two timescales are comparable close to the midplane. However, in the midplane the dead zone will likely affect the radial migration, and  grains are expected to accumulate at the boundary of the dead zone. From equation~(\ref{eqn-vertical})  we see that larger grains will settle faster than smaller grains of the same density, and different species of grains of the same size will have different settling timescales due to their density. This in accordance with the expected behaviour dictated by the stopping time  \citep{1977MNRAS.180...57W,2005A&A...443..185B}.

%-----------------------------------------------------------------------
%-----------------------------------------------------------------------

\section{Forming the bulk materials of the enstatite chondrites}
\label{sec-EHmodel}

The mineralogical properties of enstatite chondrites (ECs) differ significantly from the composition of ordinary chondrites, making them an unusual group of meteorites. Nearly pure enstatite (\ce{MgSiO3}) is the main pyroxene compound, and olivines such as forsterite (\ce{Mg2SiO4}) constitute only minor phases. Monoatomic sulfides (e.g. niningerite (MgS) and oldhamite (CaS)) are also present, together with troilite (FeS) \citep{2012Weisberg}. The minerals do not show aqueous alteration and probably did not mix with ices, while their oxygen isotopic distribution place them along the terrestrial fractionation line  \citep{2012Weisberg}. 

The ECs are divided in two sub-groups: those high in Fe (EHs) and those low in Fe (ELs), which differ in their Fe/Si and Mg/Si ratios: $0.8 \leq {\rm Fe/Si} \le 1.1$ and  $0.7\le {\rm Mg/Si} \le 0.8$ for the EHs, and $0.5\le {\rm Fe/Si} \le 0.7 $ and  $0.8\le {\rm Mg/Si} \le 0.9$ for ELs \citep{Sears97}. The EH hosts trace of refractory materials, like hibonite and melilite, which are not found in ELs \citep{2012Weisberg}. The EC assemblages  do not match the composition predicted by condensation processes when using solar elemental abundances \citep{1995GeCoA..59.3413Y,1998A&A...332.1099G}, and it has been suggested that several chemical and dynamical processes altered both their pristine bulk material and the conditions of the environment in which they formed \citep{1975GeCoA..39..735B,1979GeCoA..43.1455L}. Thus, ECs likely formed from a very unique reservoir of materials and experienced post-formation alteration.

The differences in the bulk composition between EHs and ELs  suggest that these two groups of ECs followed different pathways of formation. The presence of refractory dust  in the EHs also suggests that they formed at higher temperatures  than the ELs. 

\citet{2011Sci...333.1847N}  and \citet{2012JGRE..117.0L05W}  found that the bulk composition of the enstatite chondrites  provide good agreement with the composition and  mineralogy found on  Mercury's surface, with the exception of the iron content which is higher compared to the enstatite chondrites. \citet{2012JGRE..117.0L05W} suggest that the ECs and Mercury possibly shared the same precursor material. This would suggest that either the bulk material that formed the ECs and Mercury were co-located in the disc, or that the Solar Nebula contained two distinct zones with similar composition.

In the next two sections we discuss the possible connections between the resulting distribution of enstatite-rich condesates in our 2D disc model, the derived vertical dust settling timescales, and the chemistry of the ECs. We also introduce a simple model that suggests a research path to explain the formation of the bulk composition of ECs.

\subsection{Secondary alteration of enstatite-rich grains}
\label{subsec-alteration}

Given (1) the similarities observed between Mercury and the enstatite chondrites bulk material, (2) our finding of the two enstatite-rich zones   with similar composition (one in the optically thin surface layer and one in the midplane of our disc), (3) the presence of high refractory material on the surface of our disc, (4) the proposed common origin of the CAIs in carbonaceous and enstatite chondrites \citep{2000Sci...289.1330G}, and  (5) the  location of formation of CAIs placed in the inner upper layer of the Solar nebula \citep[and references within]{2011LPICo1639.9103M}, we suggest that the bulk material  which formed the ECs was derived from the enstatite-rich dust located in the inner surface layer of our 2D disc. 

We suggest a possible formation scenario as follows: the enstatite-rich dust (for which $fo/en\le1$) aggregated in the hot upper layer of the inner disc, accreting the refractory materials which is still abundant in the same zone (see Fig.~\ref{fig-results}). The resulting bulk material might have then vertically settled into the sulfide-rich region  where sulfidation could have occurred. 

The presence of  niningerite  in the EHs strongly supports high temperature sulfidation processes \citep{2013GeCoA.101...34L}.  \citet{2013GeCoA.101...34L} showed that sulfidation could  occur if the environment is reduced in \ce{H} content and C-rich. This can allow the presence of the necessary free \ce{S}(g) to initialise sulfidation at high temperatures. However, dissociation of  \ce{H2S}(g) can occur via other mechanisms such as  UV-photodissociation \citep{2013Chakraborty,Antonelli16122014}  during transient heating events like outbursts  and shock waves \citep{1995A&A...300L...9L}. Such high temperature events  will deplete the \ce{H2S}(g) content, enabling the formation of  \ce{S}(g) and high temperature S-bearing gas such as \ce{HS}(g) and  \ce{SiS}(g) \citep{2005Icar..175....1P} or simply dissociate  \ce{H2S}(g) \citep{1994woiki}.  However, experimental studies suggest that sulfurisation could also occur without dissociating \ce{H2S}(g) via  \ce{H2S}(g)-dust surface reaction \citep{1998M&PS...33..821L}.

High concentration of sulfur and sulfidation processes could also have contributed to the enhancement of moderate volatiles such as selenium (Se) and tellurium (Te) in bulk enstatite chondrites \citep{o1998composition,2003Treatise}.  Se and Te are chalcophiles and can be treated as a contaminant during the formation of sulfides \citep{2003ApJ...591.1220L,2010ppc..conf..347F}. However, there are several other proposed processes that can account for the abundances of Se and Te in enstatite chondrites, such as partial vaporization, incomplete condensation, weathering, secondary oxidation of sulfides, sulfur loss, and highly reducing conditions \citep{2015GeCoA.161..166K}.

Enstatite-rich grains preferentially found in lower temperature zones appear not to have accreted any (or quantitatively less) refractory material  with subsequent weak sulfidation processes occurring at lower temperatures and resulting in the ELs.

\subsection{A toy model for the formation of ECs}
\label{subsec-dustremoval}

The efficiency of the vertical settling of the dust also suggests another mechanism for the formation of ECs.  As we showed in Fig.~\ref{fig-timescale02}, vertical  settling of the dust can be an  efficient mechanism of dust sorting, since it is a function of grain density and size (see equation~\ref{eqn-vertical}). Chemical fractionation occured in the early stage of our Solar System's formation as seen from analysis of chondrites, and physical fractionation (involving dynamical processes) is one of the possible mechanisms to explain it \citep{2005mcp..book..143S}.

Vertical settling and removal of dust can lead to different chemical products which are not predicted by the classical condensation sequence in which the total bulk (dust and gas) composition does not change with time \citep{1995GeCoA..59.3413Y}. The idea of multiple separated steps which lead to the bulk composition of enstatite chondrites is not new. \citet{2000M&PS...35..601H}  investigated the condensation sequence due to  partially removing high temperature condensates and  gas (water vapour). \citet{2009M&PS...44..531B} investigated  the formation of  chondrules in enstatite chondrites via  removal of condensing liquid  droplets, which changes the chemical composition of the surrounding environment.

Here we present a toy model of gas fractionation due to dust settling and removal by the different condensation temperatures and densities of compounds.

Before proceeding, we stress the limitations and assumptions of the following calculations. As a first approximation we assume that the condensing solids are {\it totally} removed from the environment in which they condense and, once formed, they do not react with the environment. The kinetics of the condensing dust is not taken into account. Equation~\ref{eqn-vertical} applied to the  \citet{1998ApJ...500..411D,1999ApJ...527..893D} model, for a silicate grain of $a_{p}$=0.1~$\mu$m, returns a settling timescale in the order of $\tau_{set}~10^{-5}$~yr on the surface of the disc (less than one hour). This timescale will likely not allow the grain to equilibrate with the surrounding gas. The separation of a condensing particle from the gas is, thus, very efficient. As settling proceeds, the local dust density increases with settling towards the midplane and, as a consequence, grains will settle more slowly. This is clearly evident in Fig.\ref{fig-timescale02}. However, this is the case of non-growing grain, as grain growth will contribute to further decrease  the settling timescales (see eq.~\ref{eqn-vertical}).

For a gas of solar composition, at a standard pressure of $P=10^{-3}$~bar, the first major solid that condenses is iron, then forsterite followed by enstatite with decreasing  temperature \citep{1995GeCoA..59.3413Y,1998A&A...332.1099G}. ECs are fractionated in their Fe/Si and Mg/Si ratios compared to solar values. As such, iron and magnesium removal is required. Iron, for example, has a density almost twice that of forsterite so it is not unreasonable to assume that some fraction of iron grains might leave the condensation location more rapidly than forsterite. 

\begin{table*}							
\centering							
\caption[Gas abundances fractionated sequences]{Gas abundance for a solar composition ($S$) and at two subsequent fractionation cutoff temperatures. The initial solar gas mixture starts at $T_0=1850$~K. The $T_1$ gas shows the elemental abundances of  the gas at the condensation temperature of iron, $T_1=1433$~K, and the $T_2$ gas shows the elemental abundances of the  gas at the condensation temperature of forsterite, $T_2=1380$~K. The pressure is fixed at $P=10^{-3}$~bar. The bottom rows show the Mg/Si and Fe/Si values of the gas and then the same values normalised to solar. }			
\begin{tabular}	{l c c c}					
\hline	
Gas mixture	&	$S$	&	$T_1$	&		 $T_2$ \\							
(temperature)	&	($T_0=1850$~K)	&	(1433~K) &		(1380~K) \\		
\hline							
 Element	& \multicolumn{3}{c}{Abundance	(kmol)}	\\
\hline							
Al	&	$2.59 \times 10^{-4}$	&	 $1.80\times 10^{-9}$  	&	\bf $1.55\times 10^{-10}$		\\
Ar	&	$2.31 \times 10^{-4}$	&	$2.31 \times 10^{-4}$		&	$2.31 \times 10^{-4}$		\\
C	&	$2.47 \times 10^{-2}$	&	$2.47 \times 10^{-2}$		&	$2.47 \times 10^{-2}$		\\
Ca	&	$2.01 \times 10^{-4}$	&	$1.79 \times 10^{-7}$		&	$2.15 \times 10^{-8}$		\\
Fe	&	$2.91 \times 10^{-3}$	&	$1.69 \times 10^{-3}$		&	$5.91 \times 10^{-4}$		\\
H	&	92	&	92	&	92		\\
He	&	7.83	&	7.83		&	7.83		\\
Mg	&	$3.66 \times 10^{-3}$	&	$3.47 \times 10^{-3}$		&	$1.20 \times 10^{-3}$		\\
N	&	$6.22 \times 10^{-3}$	&	$3.11 \times 10^{-3}$		&	$3.11 \times 10^{-3}$		\\
Na	&	$1.60 \times 10^{-4}$	&	$1.60 \times 10^{-4}$		&	$1.60 \times 10^{-4}$		\\
Ne	&	$7.83 \times 10^{-3}$	&	$7.83 \times 10^{-3}$		&	$7.83 \times 10^{-3}$		\\
Ni	&	$1.52 \times 10^{-4}$	&	$3.18 \times 10^{-5}$		&	$3.18 \times 10^{-5}$		\\
O	&	$4.50 \times 10^{-2}$	&	$4.30 \times 10^{-2}$		&	$3.84 \times 10^{-2}$		\\
S	&	$1.21 \times 10^{-3}$	&	$1.21 \times 10^{-3}$		&	$1.21 \times 10^{-3}$		\\
Si	&	$2.97 \times 10^{-3}$	&	$2.62 \times 10^{-3}$		&	$1.47 \times 10^{-3}$		\\
\hline	
(Mg/Si)	&	1.23	&	1.32		&	0.81		\\
(Fe/Si)	&	0.98	&	0.64		&	0.40		\\
\hline									
(Mg/Si)/(Mg$_\odot$/Si$_\odot$)	&	1	&	1.07		&	0.66		\\
(Fe/Si)/(Fe$_\odot$/Si$_\odot$)	&	1	&	0.65		&	0.41		\\

\hline							
\end{tabular}							
\label{tab-abundance}							
\end{table*}

As a tentative, exploratory path for future,  more quantitative, studies we start our calculation with a high-temperature gas mixture at $T_0=1850$~K, with solar composition as reported in Table~\ref{tab-abundance}, column S. The pressure is kept constant at $P=10^{-3}$~bar. As such, we are not subscribing to any specific location of the 2D disc but instead are investigating a mechanism which can be applied to our 2D disc model\footnote{In the D'Alessio disc model, such a high pressure is only found in the midplane very close to the young Sun. Lowering the pressure results in the condensation temperature moving towards lower values, and therefore the same condensates can be found in both high pressure$+$temperature zones and lower pressure$+$temperature zones, with some possible exceptions which we will discuss in section~5.3.}.

\begin{figure*}
\centering
{\includegraphics[width=1.5\columnwidth]{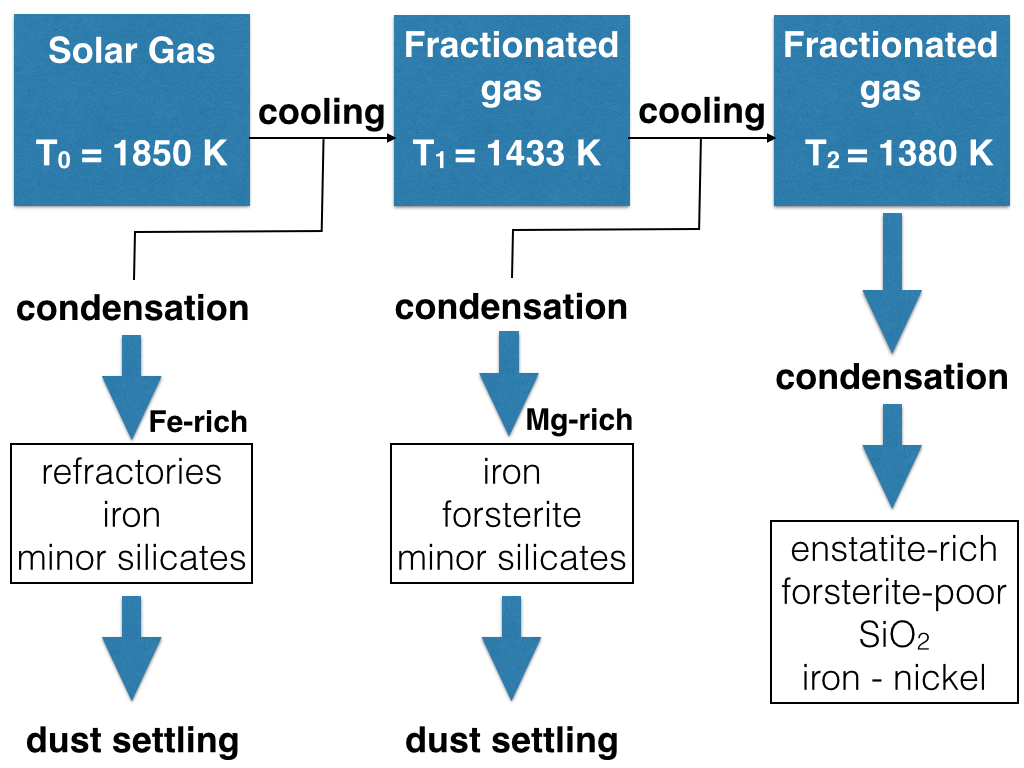}} \\
\caption[]{Schematic summarising the gas fractionation due to the vertical settling of dust. A parcel of gas with solar composition (at constant pressure of $P=10^{-3}$~bar and initial temperature of $T_0=1850$~K) is cooled to $T_1=1433$~K and all the solids which condense at this temperature are removed from the environment due to the efficient vertical dust settling. The fractionated  $T_1$ gas is then further cooled to $T_2=1380$~K, were it reaches the condensation temperature of forsterite. Forsterite, iron and minor silicates are efficiently removed from the environment, and the resulting fractionated gas has  Mg/Si and Fe/Si ratios of 0.66 and 0.41 respectively. Further condensation of the $T_2$ gas leads to enstatite-rich condensates.} 
\label{fig-secondprocess}
\end{figure*}

Since enstatite chondrites are fractionated in  Fe/Si and Mg/Si,  iron and then magnesium have to be removed from the gas. Figure~\ref{fig-secondprocess} summarises our toy model: we let the gas cool down until it reaches the condensation temperature of iron, $T_1=1433$~K, and we assume that the metal and other high-temperature solids which condense at this temperature are removed from the environment. The composition of the resulting fractionated gas is shown in Table~\ref{tab-abundance},  column $T_1$, and results in Mg/Si = 1.07, Fe/Si= 0.66 (with all values normalised to solar), while the Mg/Si and Fe/Si ratios of the removed solids are 0.43 and 3.6 respectively.

Next we let the $T_1$ gas cool further to the temperature at which forsterite condenses, $T_2=1380$~K. At this temperature, forsterite, iron and minor silicates  condense and leave the environment due to efficient vertical settling. The resulting  Mg/Si and Fe/Si  ratios of the  gas are 0.66 and 0.41 respectively (with  elemental abundances reported in Table~\ref{tab-abundance}, column $T_2$), while the Mg/Si and Fe/Si  ratios of the removed solids are 1.6 and 0.98 respectively.

A condensation sequence starting with the fractionated $T_2$ gas, at the fixed pressure of $10^{-3}$~bar, returns  enstatite, forsterite, \ce{SiO2}, iron and nickel. In general, Mg/Si ratios lower than solar lead to enstatite-rich dust \citep{2001A&A...371..133F}. The bulk composition of this dust is compatible with the bulk material found in the enstatite chondrites  \citep{2012Weisberg}.

\begin{figure*}
\centering
{\includegraphics[width=1.5\columnwidth]{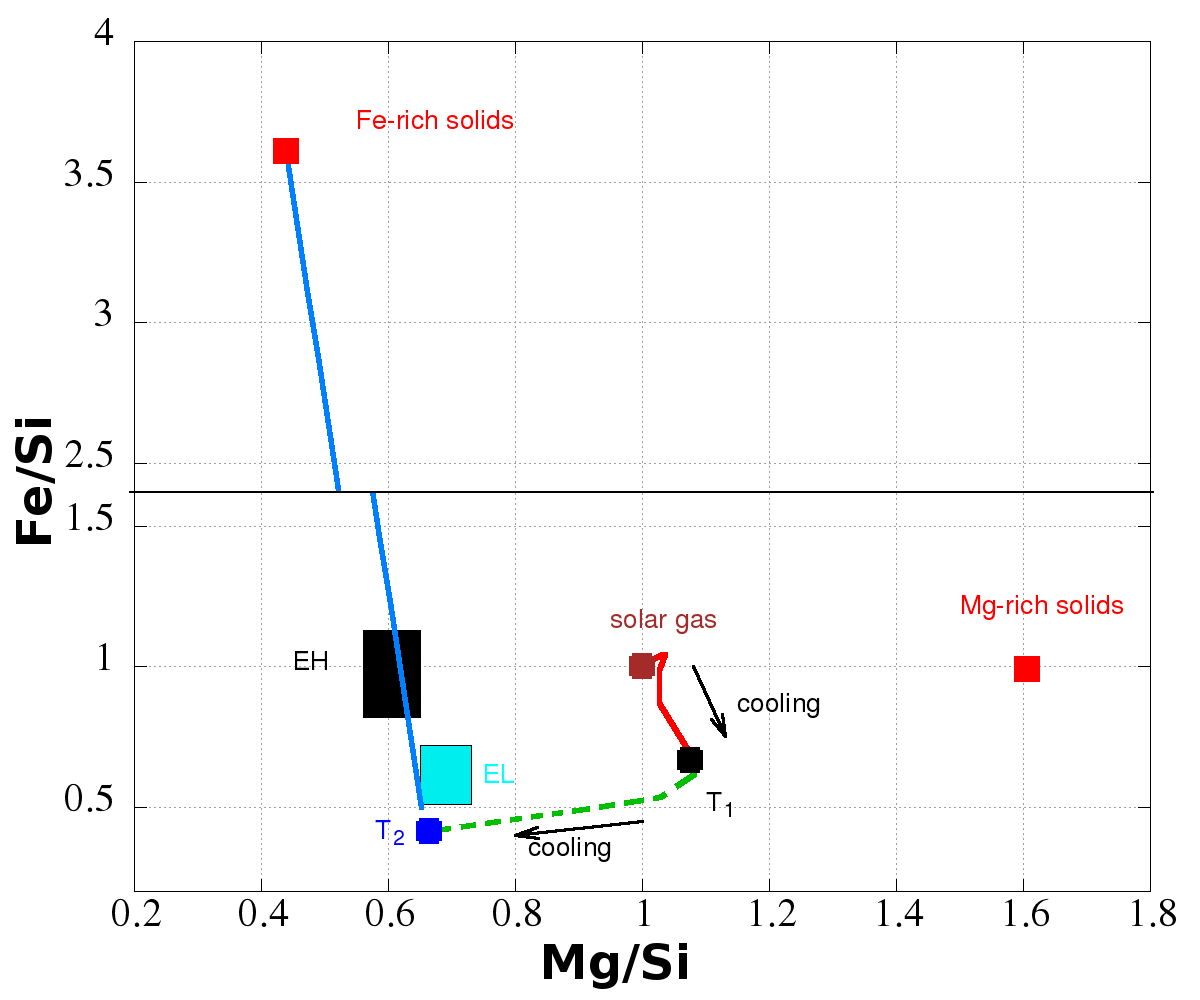}} \\
\caption[]{ Resulting Mg/Si vs Fe/Si ratios normalised to solar values from our toy model for: the initial solar gas mix, the EHs and ELs from \citet{Sears97}, the $T_1$ and $T_2$ fractionated gas from  Table~\ref{tab-abundance},  the Fe-rich solids removed after the first condensation from $T_0$ to $T_1$, and the Mg-rich solids removed after the second condensation from  $T_1$ to $T_2$. The mixing line between the $T_2$ gas and the Fe-rich solids is also drawn. The red continuous line shows how the gas with solar composition would fractionate across the temperature range from $T_0$ (solar) and $T_1$, while the green dashed line shows how the $T_1$ gas would fractionate from $T_1$ to $T_2$.} 
\label{fig-ratios}
\end{figure*}

Figure~\ref{fig-ratios} shows the Mg/Si vs Fe/Si ratios for the EHs and ELs  from \citet{Sears97},  and for the fractionated gas and condensed solids derived from our calculations, all normalised to solar. The red continuous line shows how the gas with solar composition would fractionate across the temperature range between $T_0$ and $T_1$. The green dashed line shows how the $T_1$ gas would fractionate over the temperature range between  $T_1$ and $T_2$. The $T_2$ gas is shown by a blue square and is the point at which the cooling gas reaches the forsterite condensation temperature, $T_2=1388$~K. In Fig.~\ref{fig-ratios} we also plot the Mg/Si and Fe/Si ratios of the Fe-rich grains that condensed at $T_1$ and were removed from the gas (i.e they are assumed to have settled vertically) and the Mg-rich grains condense at $T_2$ that also vertically settled.

\section{Discussion} 
\label{sec-discussion}

In this section we present the theoretical and observational evidence which supports the results of our 2D disc model. Since chondrites formed early during the protoplanetary disc phase, there might have been a connection between the chondritic bulk material forming in the inner region of the Solar Nebula, the dust present (and in theory detectable) on its surface, and the dust in the midplane, which likely formed planetesimals and planets.

Thus, in this section, we attempt to link the silicate distribution of our disc presented in section~\ref{sec-results} and the ECs bulk material described in section  section~\ref{sec-EHmodel}, with recent evidence from the Messenger observations of Mercury  and infrared observations of protoplanetary discs. Furthermore, we discuss in more detail the consequences and limitations of our proposed alternative scenario for the formation of the EC's bulk material.

\subsection{Connecting enstatite-rich dust and Mercury}
\label{connecting}

Observations and analysis from the Messenger X-Ray Spectrometer suggest that the surface of Mercury comprises Mg-rich minerals like enstatite and it is enriched in sulfur \citep{2012JGRE..117.0L05W}. 
  
1D condensation sequences have provided the theoretical background for  chemical analysis of the bulk composition of the Solar Nebula \citep{1995GeCoA..59.3413Y,1998A&A...332.1099G,2005Icar..175....1P}. Results of these 1D condensation sequences are in a general agreement with the chemical gradient found in the planets of our Solar System, when starting with an initial solar gas mixture: refractory materials at high temperatures ($T\ge1600$~K), iron and silicates at intermediate temperatures ($650\le T \rm{(K)}\le1600$), iron oxides, sulfides and water ice at lower temperatures  ($T\le650$~K). However, 1D condensation sequences can simulate only one layer of the disc at time and they cannot account for the global chemistry of discs with its multiple environments and the variegate composition of the rocky planets.

In the following discussion, we define the ``midplane'' as the vertical section of our 2D disc between $0\le Z\rm{(AU)} \le 0.02$. This vertical section represents 1/45 of the total extension of our disc model ($0.1-1.0$~AU). Looking at the 2D condensate distribution, the limitations of 1D condensation sequences clearly emerge. 

The higher temperature region in the midplane which contains iron, nickel and enstatite is confined to the inner 0.4~AU. Beyond 0.4~AU, outside the equilibrium zone, the dust chemistry cannot be predicted by equilibrium calculation. The inner enstatite-rich region in the midplane shows similarities with the enstatite-rich zone found on the surface of our disc. However, the zone for which $fo/en<1$ spans a smaller area within 0.2~AU, but the average dust composition in the zones where $fo/en<1$ and where $fo/en>1$ is the same as the respective zones present on the disc surface. 

The enstatite-rich zone  is also surrounded by a sulfide-rich zone (see Fig.~\ref{fig-results}), and given the lower temperatures, by amorphous dust. Here, kinetics become important in determining the reliability of this result. \citet{1996Icar..122..288L} studied the reaction rates of iron sulfides production via \ce{Fe} in a \ce{H2S}(g)-\ce{H20}(g) gas mixture under solar nebula conditions. They found that the timescales of \ce{FeS} production are much smaller ($\sim200$~yr) than the nebula lifetime if metal iron is present. As such, the presence of iron sulfide in this region should not be excluded.

The detection of sodium in the thin atmosphere of Mercury and the presence of moderately volatile compounds on its surface \citep{1985Sci...229..651P,2015Icar..248..547C} raise further questions regarding the presence of low-temperature material in the inner disc regions, as the amount of volatiles in the solid phase should be close to zero if the planet accreted material in the high temperature zone.

\begin{figure*}
\centering
{\includegraphics[width=1\columnwidth]{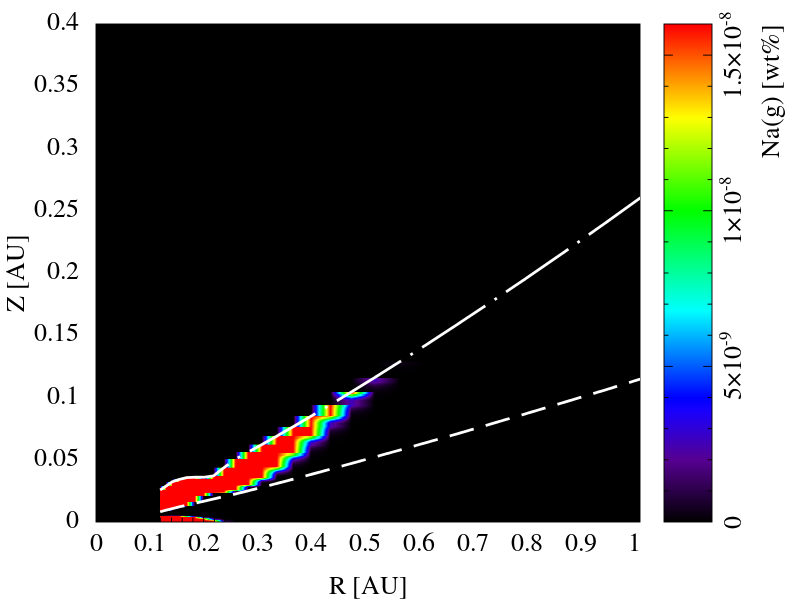}} 
{\includegraphics[width=1\columnwidth]{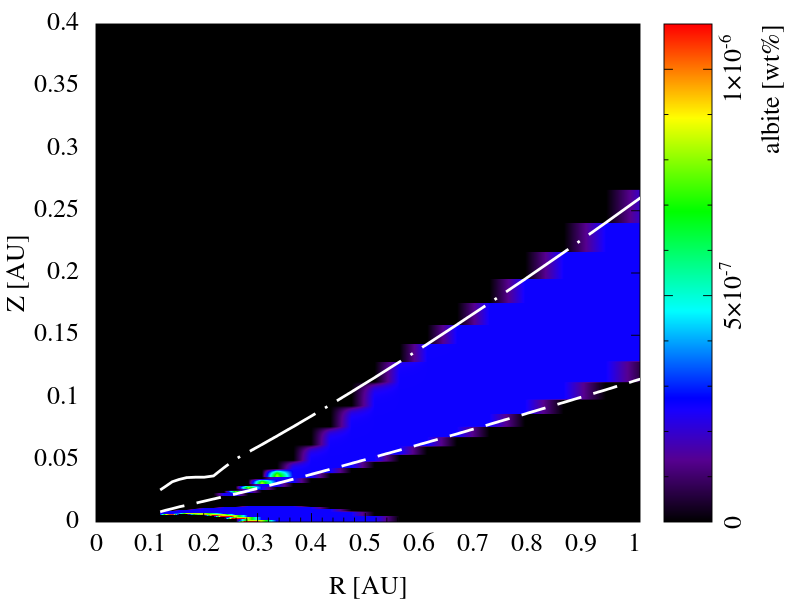}} 
\caption{Resulting chemical distribution from our 2D model for Na(g), left, and albite (\ce{NaAlSi3O8}), right. The dashed line is the $\tau=1$ surface of the disc and the dashed-dot line represents the disc surface.}
\label{sodium}
\end{figure*}

To investigate the distribution of moderately volatile material in the midplane, we report in Fig.~\ref{sodium}, as an example, the distribution of Na(g) and albite  (\ce{NaAlSi3O8}). Albite is the major Na-bearing compound at the condensation point of Na(g) according to thermodynamic equilibrium.
The enstatite-rich zone where $fo<en$ in the midplane is also Na(g)-rich. Albite is  present together with enstatite (see Fig.~\ref{fig-results}) in the zone where $fo>en$.  We thus find a distribution similar to FeS: the stability zone of albite surrounds the high temperature region where Na is in the gaseous form. However, albite condenses at a higher temperature than FeS. At a pressure of $P=10^{-3}$~bar the condensation temperature of albite is $T=970-980$~K \citep{{2011MNRAS.414.2386P}}. Thus, looking at Fig.~\ref{sodium} we see that albite becomes stable in the midplane where  $R\sim0.3$~AU. Moreover, a stability zone of sodium-rich dust is also present in the inner hotter region.

In conclusion, the midplane region of the disc up to 0.4~AU is a chemically-variegate zone in which high-temperature crystalline dust, enstatite, forsterite, metal-rich grains, processed material, sulfides, {\bf volatiles}, and unprocessed material can coexist. The presence of the dead zone in the midplane (see Fig.~\ref{fig-timescale02}) might prevent this mixture from leaving this zone.

%%%
Our resulting 2D chemical distribution with two enstatite-rich zones (disc surface and midplane) is compatible with the scenario of two distinct zones with similar bulk composition. We therefore suggest that the bulk composition of ECs could have formed from the enstatite-rich dust on the surface via the mechanism described in section~4.1, and that Mercury accreted from the enstatite-rich dust in the midplane.

Since grain growth in the disc midplane is very efficient due to the high density, with grains reaching cm-size in few thousand years \citep{2008A&A...487..265L}, accretion of this mixture in the dead zone (see Fig.\ref{fig-timescale02}), could produce enstatite-rich planetesimal on short timescales,  and the presence of low temperature regions in the inner disc, could then account for the presence of sulfides and moderate volatiles in the accreting dust.

\subsection{Silicates distribution in the disc surface}
\label{observationalevidences}

Infrared observations of the upper layers of protoplanetary discs  show a spatial variation of the  forsterite and enstatite distribution, with more enstatite  in the warm inner regions  than in the cooler outer regions where forsterite dominates \citep[]{ 2006ApJ...639..275K,2008ApJ...683..479B, 2009A&A...497..379M}. The reason for this distribution remains uncertain. 

 \citet{2007ApJ...659..680K} investigated the possible disc location from which the $10~\mu$m silicate feature could arise. The 10~$\mu$m silicate feature is usually modelled  assuming a larger contribution of enstatite  than forsterite \citep[]{2006ApJ...639..275K,2008ApJ...683..479B,2009A&A...497..379M}.  \citet{2007ApJ...659..680K} derived a relation  which links the distance from the protostar where the 10~$\mu$m feature arises, $R_{10}$, with the luminosity of the star, $L_{\star}$,  given by
\begin{equation}
\label{10microns}
R_{10}= 0.35~{\rm AU} \times \Bigl[\frac{L_{\star}}{L_{\odot}}\Bigr]^{0.56} \,.
\end{equation}
Applying equation~(\ref{10microns}) to our disc model, we find $R_{10}$= 0.52~AU, which is in the region where enstatite and forsterite coexist in the surface layer of our disc (see Fig.~\ref{fig-results}). 

The region of the discs for which $fo/en<1$ ($\sim$0.2-0.5~AU) in our resulting chemical distribution is slightly offset compared to the calculated value of 0.52~AU. This difference may be due to the different grain size distribution used to model the opacity in the disc. We use the well-mixed model with  $a_{\rm min}=0.005~\mu$m and $a_{\rm max}=1.0$~mm  \citep{1998ApJ...500..411D,1999ApJ...527..893D}, while  $a=0.1~\mu$m for \citet{2007ApJ...659..680K}. The smaller grains in  \citet{2007ApJ...659..680K} model increase the temperature in the disc moving the location of the 10~$\mu$m feature toward larger radii.  

Thus, our derived 2D distribution of silicates suggests that the enstatite-rich dust observed in the $10~\mu$m feature can be of condensation or thermal annealing origin. Furthermore, we suggest that the dust composition derived in section~\ref{sec-results} could be used  as a raw model to characterize the crystalline component of the dust necessary to fit the 10~$\mu$m feature of the infrared spectra of protoplanetary discs.
 
\subsubsection{The forsterite problem}
\label{forsteriteproblem}
 To account for the forsterite observed in the outer regions of discs, \citet{2000A&A...364..282F} and \citet{2002ApJ...565L.109H} suggest that thermal annealing of enstatite from heating shocks occurs in the outer part of the disc. Other theories point to an efficient radial grain transport mechanism which distributes the forsterite  formed via condensation in the inner zones of the disc  toward the cooler regions \citep{2004Natur.432..479V,2012ApJ...744..118J}. Our  2D distribution of enstatite-rich dust is in good agreement with both theories since it can constitute the bulk material from which forsterite can form.

However, in our calculations we see that the abundance of forsterite increases with increasing stellar distance, while the abundance of enstatite decreases. In the region of the surface of the disc between $0.5\le\rm{R (AU)}\le0.8$ we find $fo/en>1$. If we consider the entire zone in which forsterite coexists with enstatite, the average dust composition results in   forsterite (\ce{Mg2SiO4}) 24.9~wt\%, enstatite (\ce{MgSiO3}) 19.8~wt\%, metals (\ce{Fe}-\ce{Ni}) 43.1~wt\%, troilite (\ce{FeS}) 9.6~wt\% and others 2.6~wt\%. 
This resulting chemistry is also compatible with infrared observation which show that forsterite dominates between  $\lambda\sim20-30~\mu$m  \citep{2008ApJ...683..479B}, where temperatures are cooler than the $\lambda\sim10~\mu$m region. 

This general agreement between $20-30~\mu$m observation and our equilibrium calculations poses a new question. As pointed out in section~\ref{subsec-disco},  the assumption of equilibrium ceases to be valid when moving toward lower temperatures (increasing distance from the star). The presence of forsterite at $T\le1000$~K for an initial gas mixture with a solar composition is predicted by thermodynamic calculation \citep[and references within]{2011MNRAS.414.2386P}, but is controversial since kinetics can prevent its formation \citep{1995GeCoA..59.3413Y,1998A&A...332.1099G}. Experiment studies on annealing of amorphous grains found that amorphous enstatite is converted to crystalline forsterite at $T\sim1000$~K \citep{2000ESASP.456..347F,2011A&A...529A.111R}. However, this good agreement between observation and our 2D calculations suggests the utility of further experimental studies on the condensation, crystallisation and thermal annealing of silicates grains at temperature $T\le1000$~K.

\subsection{Dust fractionation and sedimentation}
\label{fracsedim}
An alternative scenario for the formation of the ECs bulk material is the fractionation and sedimentation of the condensates as discussed in section~4.2. A problem with the model we presented is that we started with $P\sim10^{-3}$~bar, and such high pressures are only found in the very inner midplane of the D'Alessio disc model that we used. For fractionation to occur, we ideally need to start at the surface of the disc where the dust settling is very efficient, but the pressure is much lower.

It has to be noted that at very low pressure, thermodynamic calculations predict that the condensation of forsterite  occurs at similar (or higher) temperature relative to iron \citep{1990E&PSL.101..180P,2005Icar..175....1P,2011MNRAS.414.2386P}. This would clearly affect the sequence of condensed material proposed in section~4.2. Following a qualitative prediction, we can expect first the fractionation of magnesium  of the gas, followed by fractionation of the iron. In this case, the fractionated gas will first move toward the EHs, lowering its Mg/Si ratio, and then toward the ELs with low Fe/Si ratio -- which on Fig.~\ref{fig-ratios} would move the solar gas left towards the EHs material and then down towards the ELs material. An investigation of the fractionation of gas at different pressures will be helpful to further determine the location in the disc in which different fractionation can occur.  In either case, both fractionation pathways lead to the ECs zone of the Fe/Si and Mg/Si ratios. The resulting enstatite-\ce{SiO2}-rich assemblage could then vertically settle towards the sulfur-rich region.

Thus all the bulk material necessary to form of ECs is present in this zone of the disc and a working hypothesis is that the ECs could be the final products of grain sorting followed by sulfidation. Although the fractionation of the refractory  dust occurs at  high temperatures, traces of Ca(g) and Al(g) are preserved during the fractionation process. This could account for the presence of Ca-Al compounds in the ELs. It is interesting to note that according \citet{2012Weisberg} (their table 2), there is no evidence in EL of high refractory material such as corundum, hibonite or melilite, which in our calculation are indeed removed in the first step.

However, it can also be seen that a mixing line between  Fe-rich grains  and  $T_2$ gas crosses the EHs (Fig.~\ref{fig-ratios}). This suggests that together with the hypothesis that EHs and ELs formed in different location,  the re-introduction of mostly Fe-rich dust (from various transportation mechanisms) into fractionated gas and dust might also account for the differences found in EHs and ELs.

The calculated vertical settling timescales are also consistent with Spitzer observations of crystallisation processes occurring in the young star Ex Lupi \citep{2009Natur.459..224A}: forsterite grains were produced on the surface layer of the disc  from amorphous silicate dust  due to the high temperatures developed during stellar outbursts. The observed forsterite grains then become undetected within months suggesting that rapid transport mechanisms of grains (vertically and/or radial) take place in the disc \citep{2009Natur.459..224A,2012ApJ...744..118J}.

%-----------------------------------------------------------------------
%-----------------------------------------------------------------------

\section{Conclusions}
\label{sec-conclusions}

In this work we  presented, for the first time, a 2D distribution of condensates  within the solar nebula.  We consider the optically thin region of the disc surface within 0.8~AU  from the Sun and the optically thick disc midplane out to 0.4~AU, where equilibrium modelling is feasible in terms of chemical reaction rates. 

The resulting distribution revealed a complex chemistry: we found two enstatite-rich zones in the disc with similar bulk composition: one in the inner midplane of the disc and one on the surface layer. We present models coupling the chemistry and dynamics that support the idea that the bulk material of ECs and Mercury formed in these two distinct regions from material of similar bulk composition. 

Our distribution of forsterite and enstatite in the surface layers of the disk appears to be compatible with infrared observations of the silicate distribution in protoplanetary discs. We see enstatite-rich dust in the inner region of the disc surface which is likely the source of the $10~\mu m$ feature observed in protoplanetary discs. We suggest a dust composition which can be used to characterize the crystalline component of the dust necessary to fit the 10~$\mu$m feature of the infrared spectra of protoplanetary discs.

In conclusion, our results show that the inner Solar Nebula was  an enstatite-rich environment and that there is a possible link between ECs, Mercury and the enstatite-rich dust in protoplanetary discs observed in the infrared.

Vertical dust settling is an efficient process in discs, and we showed that condensation sequences fractionated by size and species sorting may also result in a bulk material consistent with the Mg/Si and  Fe/Si ratios of the ECs. However, more detailed and quantitative models are required to explore the validity of our model.

Given the complexity of coupling chemistry, kinetics and dynamics in protoplanetary disc models, this work represents only a preliminary study of the many possible pathways to the ECs and rocky planets.

\section{Acknowledgments}
\label{acknowledgments}

This work was partially supported by the Swinburne University Postgraduate Research Awards (FCP),  CSIRO Astrophysics and Space Science (KL) and the visiting professor scheme from Universit\'e Claude Bernard Lyon 1 (STM). FCP is grateful to the LABEX Lyon Institute of Origins (ANR-10-LABX-0066) of the Universit\'e de Lyon for its financial support within the program "Investissements d'Avenir" (ANR-11-IDEX-0007) of the French government operated by the National Research Agency (ANR).

The authors wish to thank the late Paola D'Alessio for providing  the detailed structure of the disc model, Jean-Fran\c{c}ois Gonzalez for helping with the visualisation of the data, and the anonymous referee for suggesting further investigations into the distribution of moderately volatile materials which we hope has improved the manuscript.

%---------------------------------------------------------------------

%---------------------------------------------------------------------

\newpage

\bsp

\label{lastpage}

\end{document}